\documentclass{cernyrep}
\usepackage[T1]{fontenc}
\usepackage[bookmarks, colorlinks=true, linktoc=page, pdftex, linkcolor=black, citecolor=black, urlcolor=blue]{hyperref}
\usepackage{fancyhdr}
\fancyhfoffset{4 mm}
\fancypagestyle{ARTTITLE}{%
\fancyhf{} 
\lhead{\hfill CERN Accelerator School Proceedings ---
{\it RF for Accelerators} ---  Berlin, Germany,  2023\hfill}
\lfoot{\hspace{3mm} Available online at \url{https://cas.web.cern.ch/previous-schools}}
\rfoot{\thepage\hspace*{3mm}}
 
}

\usepackage{varwidth}
\usepackage{xcolor}
\begin{document}

\title{Matching and Power Coupling}
\author{Graeme Burt}
\institute{The Cockcroft Institute, Lancaster University, UK}

\begin{abstract}
A key factor in any RF system is the mechanism for coupling the RF power from an amplifier into an accelerating cavity. Any tranmission line will experience reflections if there is a mismatch in the impedance between the line and its load. In accelerating cavities due to their high quality factors there is often a large mismatch between the cavity shunt impedance and the tranmission line. This lecture will look at how to overcome this mismatch and ensure efficient coupling without steady-state reflections in both standing-wave and travelling-wave cavties.
\end{abstract}
\maketitle
\thispagestyle{ARTTITLE}

\section{Introduction}
RF is coupled into and out of the RF cavity from a waveguide or any other transmission line. The~coupler can come in rectangular waveguide or coaxial configurations. The interface between the cavity and the~coupler can couple via the electric fields, magnetic fields or both. At the interface the incoming wave can be transmitted into the cavity or reflected back into the transmission line. The process of maximising the~transmission into the cavity and minimising reflections is known as matching.
\section{RF Couplers}\index{Mode matching}
At an interface between two RF sections the fields must be continuous at the interface to satisfy Maxwell’s equations, ie the electric and magnetic fields must be identical on each side of the interface. In addition to this the boundary conditions on any metallic walls must be preserved, $E_{||}=H_t=0$. If the two RF sections have different cross sections it is not possible to match a single mode in one section to a single mode in the other section and coupling into multiple mode occurs. In the case of waveguide couplers, the field in the waveguide mode (normally the $\textrm{TE}_{10}$ mode) should be matched to the fields in the cavity, with electric and/or magnetic fields aligned in the same direction on either side of the interface. The coupling between the electric fields can be found by matching the cavity field at the interface $\vec{E_{cav}}$ to an expansion in terms of the modes inside the coupler
\begin{equation}
    \vec{E_{cav}}= \sum_{n=1}{a_n \vec{E_{n,coup}}},
\end{equation}
where  $\vec{E_{n,coup}}$ is the electric field of the $n^{th}$ waveguide mode at the same interface and $a_n$ is the amplitude of that waveguide mode. Similarly, the magnetic field at the cavity interface, $\vec{B_{cav}}$, is expanded as
\begin{equation}
    \vec{B_{cav}}= \sum_{n=1}{a_n \vec{B_{n,coup}}},
\end{equation}
where  $\vec{B_{n, coup}}$ is the magnetic field of the $n^{th}$ waveguide mode at the interface and $a_n$ is the amplitude of that waveguide mode. This equation can be solved for each waveguide mode to find the coupling to each mode.
However the need to match both E and H simultaneously cannot be achieved with a single coefficient. At any discontinuous interface with a wave travelling towards that interface, the wave will be scattered by that interface into forwards and backwards waves. Depending on the type of discontinuity the electric fields from the forwards and backwards waves will either add or subtract, while the magnetic fields will do the opposite. This means by the inclusion of backwards wave we can match both sets of fields at the interface. The electric fields are now given by,
\begin{equation}
    \vec{E_{cav}}= \sum_{n=1}{a_n \vec{E_{n,coup}}} + \sum_{n=1}{b_n \vec{E_{n,coup}}},,
\end{equation}
where $a_n$ is the amplitude of that forwards (incoming) waveguide mode, and $b_n$ is the amplitude of the~backwards (reflected) waveguide mode. If we take an aperture connecting a cavity to a waveguide, we can use a Fourier series of sinusoidal terms in the waveguide to replicate the fields from the cavity presented to the aperture. Unless the cavity fields are a perfect half period across the aperture there will be several spacial harmonics in the waveguide. As the fields in a waveguide are discretised into discrete modes with m half-wave variations across the width and n half-wave variations across the height, each harmonic on the Fourier series represents a different waveguide mode. As we can have forwards and reverse waves this would not be single valued, but the requirement to match both E and H reduces it to a single solution comprising of a finite number of mode amplitudes. As most waveguide power couplers are designed to have a single mode above cut-off at their operating frequency all other modes will be cut-off. These cut-off modes are localised around the interface, decaying exponentially into the waveguide, but they are still important as the absorb power and locally increase fields around a coupling aperture.

\begin{figure}
\centering
\includegraphics[width=120mm]{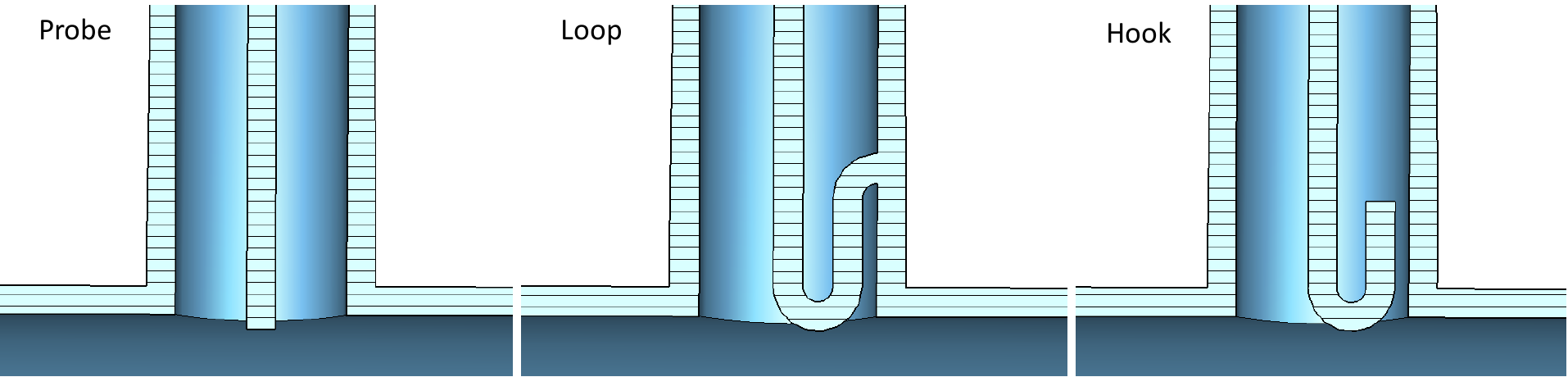}
\caption{\label{fig_Probetypes} The three types of termination for a coaxial coupler: probe, loop and hook.}
\end{figure}

\begin{figure}
\centering
\includegraphics[width=120mm]{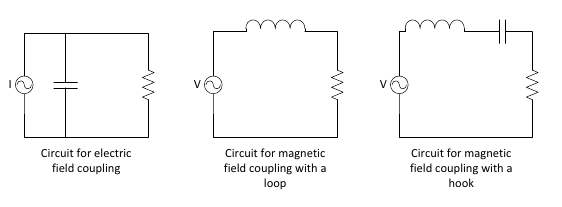}
\caption{\label{fig_HOMcouplingcircuit} Equivalent circuits for electric and magnetic coupling.}
\end{figure}

For coaxial couplers we have a choice in the geometry at the end of the coupler where the cavity and coupler meet, that we can optimise to ensure the cavity is critically coupled. If we leave the inner conductor un-terminated with no connection to the outer conductor (known as probe termination), as shown in Fig.~\ref{fig_Probetypes}, then the electric field of the cavity can create a charge difference between the inner and outer conductor which varies with time, hence acting as a current source in parallel with the capacitance between the inner and outer conductor. The current, $I$, is given by
\begin{equation}
    I=- \frac{\mathrm{d}Q}{\mathrm{d}t} = -\epsilon_0 \frac{\mathrm{d} \int_{tip}{\vec{E}.\mathrm{d}\vec{S}}}{\mathrm{d}t},
\end{equation}
where $E$ is the electric field on the tip of the~inner conductor and $S$ is the surface area of the tip of the~inner conductor.
If we connect the inner conductor to the outer conductor via a loop then the magnetic field can create a voltage across the hook loop via magnetic induction.  This has an equivalent circuit diagram of a voltage source in series with an inductor. The voltage is given by
\begin{equation}
    V=- \frac{\mathrm{d} \Phi}{\mathrm{d}t} = - \frac{\mathrm{d} \int_{loop}{\vec{B}.\mathrm{d}\vec{S}}}{\mathrm{d}t},
\end{equation}
where $\Phi$ is the magnetic flux through the loop. A magnetic loop has difficulties in assembly as the~inner and outer conductors need to be joined. It is also possible to instead have an inductive hook at the end of the coaxial lines inner conductor that has a small capacitive gap between itself and the outer conductor, also shown in Fig.~\ref{fig_Probetypes}. Such a termination can be excited by both electric and magnetic fields, however each has a slightly different equivalent circuit. For the hook, the inductor and capacitor are in series with each other, however for magnetic field coupling, this series LC circuit is also in series with the voltage source and for electric coupling, the series LC circuit is instead in parallel with the current source. As the~capacitor, $C_{gap}$, and inductor, $L_{loop}$, are in series, they form a resonant circuit which acts as a bandstop filter for electric fields and a bandpass filter for magnetic fields, with a resonant frequency
\begin{equation}
    \omega_f=\frac{1}{L_{loop} C_{gap}}.
\end{equation}
The equivalent circuit for each type of coupling is shown in Fig.~\ref{fig_HOMcouplingcircuit}. The choice between types will depend on the chosen coupler location, the cavity fields at that location, and the RF heating on the coupler tip.

\subsection{Coupling Power into an RF Structure}
To connect the RF power supply to the cavity we must construct an antenna that will radiate power into the cavity this avoids the power being reflected back up the waveguide. This is normally just a waveguide or coaxial line connected to the cavity via a small hole in the beam pipe or the cavity walls, known as an~input or power coupler, which will be discussed later. There will be a discontinuity between the cavity and the coupler with an impedance mismatch, which leads to reflections. However this is complicated as some of the power that gets into the cavity will be coupled back to the coupler. The reflected wave and the power coming from the cavity will combine together and interfere. Additionally as the stored energy in the cavity increases so will the power leaking back to the coupler hence the backwards traveling wave will vary in time until the cavity reaches steady-state.

The strength of the coupling can be represented by defining an external $Q$ factor, $Q_e$, which relates the stored energy in the cavity to the power that would flow into the coupler if there is no RF power being supplied to the cavity, $P_e$; this is given as \index{Q factor}
\begin{equation}
Q_e= \frac{\omega U}{P_e}.
\end{equation}
It is convenient to add the external power lost to the coupler with the input power turned off, $P_e$, to the~cavity ohmic losses, $P_c$ to give the total losses with the RF supply off $P_t$. Since $P_t = P_e + P_c$ then we can also define a quality factor combining all losses known as the loaded $Q$ factor, $Q_L$, where for a~cavity with a single coupler,
\begin{equation}
\frac{1}{Q_L} = \frac{1}{Q_0} +  \frac{1}{Q_e} .
\end{equation} 
It is also useful to define the coupling factor, $\beta$, which is the ratio of losses through the coupler to the~ohmic losses in the cavity walls
\begin{equation}
\beta = \frac {P_e}{P_c} = \frac{Q_0}{Q_e}.
\end{equation} 
The cavity will have a finite bandwidth over which power is coupled into the cavity. The impedance of the cavity can then be solved from the equivalent circuit as a function of frequency $\omega$ for 
\begin{equation}
    Z=\frac{R_{s,circuit}}{1+i Q_L \left( \frac{\omega}{\omega_0}-\frac{\omega_0}{\omega}.\right)}.
\end{equation}
This gives a full width half maximum bandwidth in Z of $2 Q_L/\omega$.

At frequencies outside of this bandwidth all the power will be reflected. When we have an RF pulse the rising and falling edges of the pulse will contain a wide range of frequencies, some of which will fall outside the band and will be reflected. For a square pulse, almost all of the power at the rising edge will be outside the band and all of the power will initially be reflected, but over time the bandwidth will reduce, reducing reflections and increasing the power coupled into the cavity. For slower rise times and larger cavity bandwidths the reflections are reduced. To model this, we can consider an equivalent circuit. When RF power is supplied to the cavity there will be a large impedance mismatch between the~coupler, which will have an impedance of a few tens of ohms, as it is required to have a high power flow for transport, to the cavity which will have an impedance of several M$\Omega$ in order to reach high gradients with minimal power. This means that at the interface between the coupler and the cavity there will be a large reflection in anti-phase to the supplied RF power. This reflected signal from the interface will interfere with the power leaking back into the coupler from the cavity which will be in phase with the supplied RF power. The total power flowing back up the coupler, $P_r$, when driving the cavity on resonance, will be equal to
\begin{equation}
\label{eqn_ReflectU}
P_r= \Bigg(\sqrt{P_f}-\sqrt{ \frac{\omega U}{Q_e}} \Bigg)^2,
\end{equation}
where $P_f$ is the forward power from the RF source. The reflection from the interface between the cavity and coupler due to the mismatch will be slightly less than 100\% in reality. We will refer to the total reverse power going back up the waveguide as the reflected power $P_r$, the power reflected from the~interface between the cavity and coupler when the cavity is empty as the interface reflection, $P_i$, and the~power leaking back up the coupler from the stored energy as the emitted power, $P_e$.
The change in stored energy over time in an RF cavity without beam can be obtained by summing the power flowing into and out of the system as
\begin{equation}
\frac {\mathrm{d}U}{\mathrm{d}t}= P_f-\Bigg(\sqrt{P_f}-\sqrt{ \frac{\omega U}{Q_e}} \Bigg)^2-\frac{\omega U}{Q_0}.
\end{equation}
Expanding the brackets gives
\begin{equation}
\frac {\mathrm{d}U}{\mathrm{d}t}= \sqrt{ \frac{4 P_f \omega U}{Q_e}}-\omega U \Big( \frac{1}{Q_0} +  \frac{1}{Q_e} \Big)
\end{equation} 
and inserting the definition of loaded $Q$ factor into this equation gives us
\begin{equation}
\frac {\mathrm{d}U}{\mathrm{d}t}= \sqrt{ \frac{4 P_f \omega U}{Q_e}}- \frac{\omega U}{Q_L}.
\end{equation} 
We can hence find the steady-state stored energy, $U_0$, when the stored energy no longer varies with time ($\mathrm{d}U/\mathrm{d}t=0$), by solving the quadratic equation for $\sqrt{U}$,
\begin{equation}
\label{eqn_U0}
U_0 =  \frac{4 P_f Q_L^2}{Q_e \omega}= \frac{4 P_f \beta}{(1+\beta)^2} \frac{Q_0}{\omega},
\end{equation}
This can also be rearranged to give the forward power required for a cavity to reach a given voltage
\begin{equation}
\label{eqn_Pf}
P_f =  \frac{V_c^2 (1+\beta)^2}{8 R \beta},
\end{equation}

\begin{figure}
\centering
\includegraphics[width=120mm]{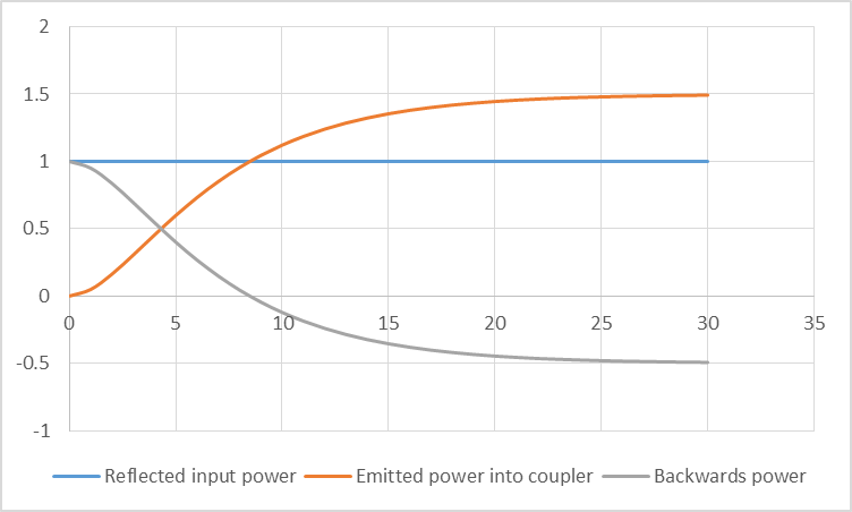}
\caption{\label{fig:superposition} The reflected, emitted and backwards travelling signals in an over-coupled cavity}
\end{figure}

We can also solve the time dependence of the stored energy, assuming the initial energy is zero, by solving the first-order nonlinear ordinary differential equation as
\begin{equation}
\label{eqn_Uvst}
U = U_0  \bigg( 1 - e^{-\omega t/2 Q_L} \bigg)^2.
\end{equation}
It can be seen that the stored energy increases with time as the cavity fills with RF energy, converging to $U_0$. The time constant for the filling, $\tau$, is
\begin{equation}
\tau=\frac{\omega}{Q_L}
\end{equation}
which is inversely proportional to the loaded $Q$ factor rather than the ohmic $Q$ factor. Having solved for the stored energy we can return to solving the reflected power; inserting Eq.~(\ref{eqn_U0}) into Eq.~(\ref{eqn_ReflectU}) we obtain the steady-state reflected power \index{reflected power}
\begin{equation}
P_r= P_f \Bigg( 1- \frac{2 Q_L}{Q_e} \Bigg)^2
\end{equation}
and inserting the definition of the coupling factor we obtain
\begin{equation}
P_r= P_f \Bigg( 1- \frac{2 \beta}{1+ \beta} \Bigg)^2. 
\end{equation}
It can be seen that when $\beta=1$ the reflected power (flowing back up the coupler) is zero and the cavity is said to be critically-coupled. This can be interpreted as the reflections from the interface -- due to the~impedance mismatch between the cavity and the waveguide -- exactly cancelling out the power emitted from the cavity into the waveguide as they will have equal magnitude but will be 180$^\circ$ out of phase. $\beta=1$ when the ohmic and external $Q$ factors, and hence the external coupler and ohmic losses, are equal. When $\beta>1$ the ohmic $Q$ factor is greater than the external $Q$ factor and hence the coupler is said to be over-coupled; when $\beta<1$ it is said to be under-coupled. This can also be rearranged to find the coupling factor by measuring the steady-state reflections from a cavity to give \index{under-coupled cavity}\index{critically-coupled cavity}\index{over-coupled cavity}
\begin{equation}
\beta= \frac{1 \pm \sqrt{P_r/P_f}}{1 \mp \sqrt{P_r/P_f}},
\end{equation}
with the upper sign used if $\beta>1$ and the lower sign used if $\beta<1$. Often $\sqrt{P_r/P_f}$ is referred to as the~input port reflection coefficient, $S_{11}$, which is the first element in the scattering matrix of reflected and transmitted waves from a multiport RF network~\cite{Pozar}. By inserting Eq.~(\ref{eqn_Uvst}) into Eq.~(\ref{eqn_ReflectU}) we can also solve for the reflected power for the case of the time-dependent reflections
\begin{equation}
P_r= P_f \left[ 1- \frac{2 \beta}{1+ \beta} \bigg( 1 - e^{-\omega t/2 Q_L} \bigg) \right]^2.
\end{equation}
The first term represents the interface reflection and the second term the emitted power from the cavity. For the case of a critically- or under-coupled cavity, the reflections are initially close to 100\% as there is no stored energy in the cavity to cancel the reflections at the interface. As the stored energy builds up in the cavity, so does the power emitted back down the coupler from the cavity, cancelling out some of the power reflected at the interface and reducing the power flowing back up the coupler. For a critically-coupled cavity the reflected power tends to zero over the filling time of the cavity, while for an under-coupled case they tend to a finite value. For an over-coupled cavity the behaviour is initially identical but soon the emitted power grows larger than the interface reflection. Hence the superposition of both reverse signals causes the reflected power to reduce to zero and then start increasing again to a finite value, as first the interface reflection dominates the reflected signal, then the emitted power. As these two signals have a 180$^\circ$ phase difference, the phase of the reflected signal also changes by 180$^\circ$ as it crosses zero. This is illustrated in Fig.~\ref{fig:superposition} showing the reflected, emitted and backwards travelling signals.

\begin{figure}
\centering
\includegraphics[width=120mm]{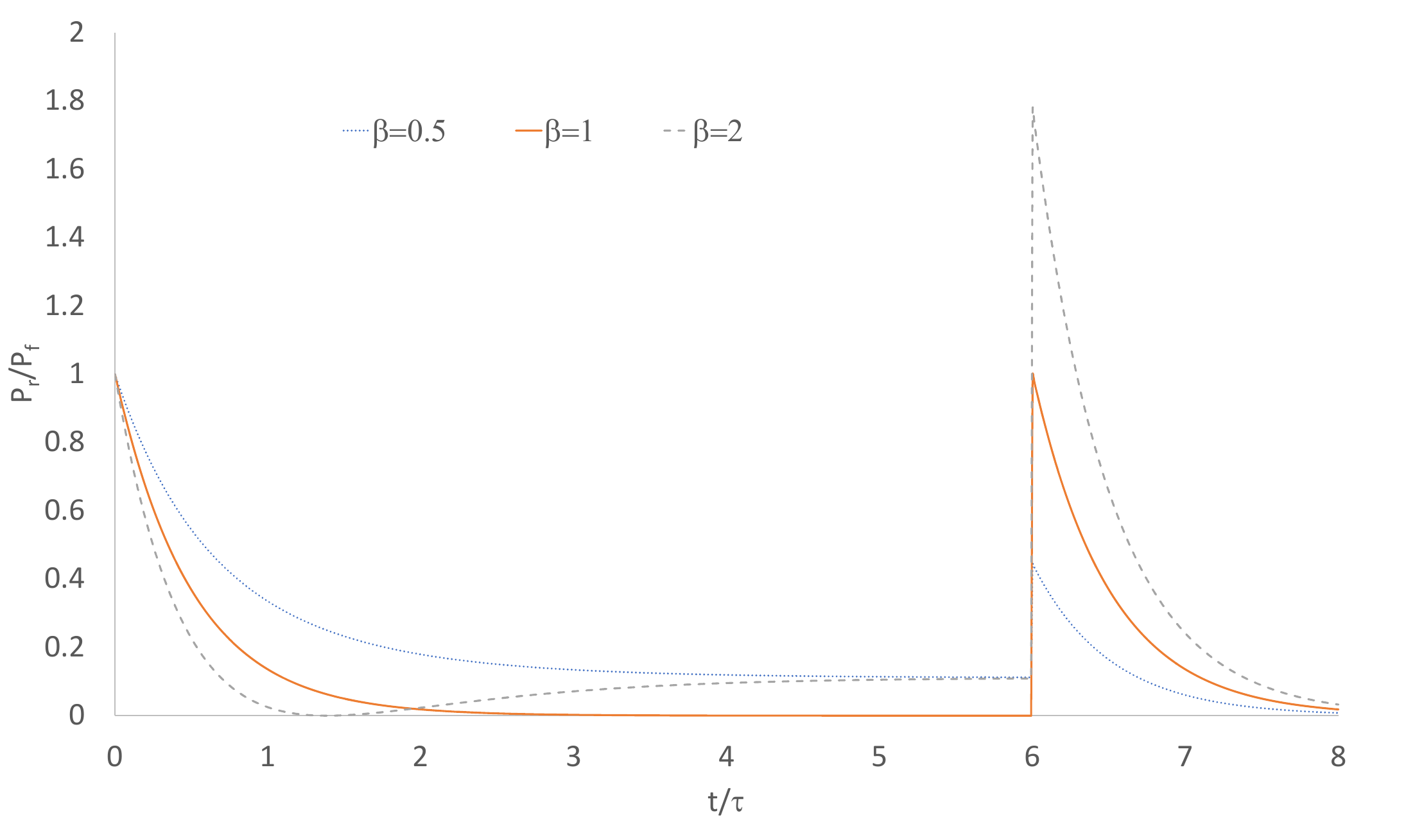}
\caption{\label{fig_ReflectTransient} Transient reflected power for a square wave input pulse, where $t_{pulse}=6\tau$ for $\beta$ = 0.5, 1 and 2.}
\end{figure}

When the RF is switched off suddenly, $P_f$ becomes zero, hence it no longer cancels the emitted power and the reflected power will again spike with the peak reflected power directly after the RF pulse is switched off given by 

\begin{equation}
P_r= \frac{\omega U}{Q_e} = P_f \Bigg(  \frac{2 \beta}{1+ \beta} \Bigg)^2
\end{equation}
with an over-coupled cavity creating a reflected power spike up to four times the power of the initial forward RF power, a critically-coupled cavity reflected power spike the same size as the forward power and the under-coupled cavity with a smaller spike than the forward power. The reflected signals from a~square envelope pulse, of duration $t_{pulse}$, for each case is shown in Fig.~\ref{fig_ReflectTransient}. The stored energy in the~cavity will decrease exponentially with the time constant of the cavity

\begin{figure}
\centering
\includegraphics[width=15cm]{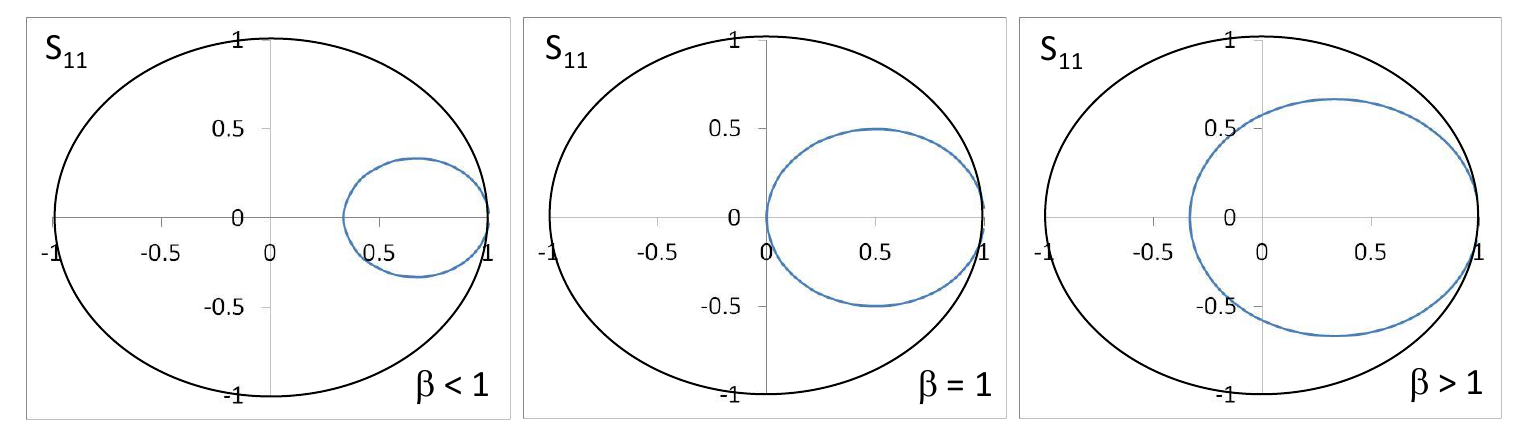}
\caption{\label{fig_ReflectPolar} The reflected signal as a function of frequency on a polar plot for $\beta$ = 0.5, 1 and 2.}
\end{figure}

\begin{equation}
U = U_0 e^{-\omega t/Q_L}.
\end{equation}
Hence the stored energy will vary with time as
\begin{equation}
U =U_0 e^{-\omega (t-t_{pulse})/ Q_L}=\frac{4 P_f Q_L^2}{Q_e \omega} e^{-\omega (t-t_{pulse})/ Q_L}.
\end{equation}
The cavity voltage can then be obtained from the $R/Q$ of the cavity. When the RF is turned off we can set the forward power to zero in Eq.~(\ref{eqn_ReflectU}) but maintaining the same stored energy at the moment the~RF is turned off; this will then decay exponentially. This yields
\begin{equation}
P_r= P_f \left[ \frac{2 \beta}{1+ \beta} \bigg( e^{-\omega (t-t_{pulse})/2 Q_L} \bigg) \right]^2,
\end{equation}
where $P_f$ is the power before the RF is turned off at time $t_{pulse}$. If the RF drive frequency, $\omega$, is different than the cavity resonant frequency, $\omega_0$, the steady-state reflected power  can be given by~\cite{Padamsee}
\begin{equation}
P_r= P_f \Bigg(  \frac{1- \beta -i Q_0 \delta}{1+ \beta+i Q_0 \delta} \Bigg)^2,
\end{equation}
where $\delta$ is given by
\begin{equation}
    \delta = \frac{\omega}{\omega_0}- \frac{\omega_0}{\omega}.
\end{equation}Plotting the reflected signal as a function of frequency on a polar plot will give a circle which will not enclose the origin if under-coupled, will cut through the origin if critically-coupled and will enclose the~origin if over-coupled; this is shown in Fig.~\ref{fig_ReflectPolar}. This allows the coupling to be measured from the~reflected power.

\subsection{Fundamental Power Couplers}\index{fundamental power coupler}
The RF is fed into the cavity via a fundamental power coupler (FPC), which is designed to handle high power flow. By varying the geometry of the coupler, hence altering their capacitance and/or inductance, we can vary the external $Q$ of the coupler, in order to match the RF systems. For high-frequency normal conducting cavities, the FPC is almost always waveguide for power handling reasons, while for low-frequency cavities (below 400 MHz) coaxial coupling is preferred to reduce the size. The couplers can be placed in the cavity equator, known as on-cell couplers, or beside the cavity to couple via the~beam pipe. SRF cavities normally prefer coaxial couplers, even at higher frequencies, to reduce the heat transport through large waveguides, although for synchrotrons where high power is required, rectangular waveguide couplers are sometimes used as it avoids the problem of cooling the inner conductor. The~presence of a coupling hole near the cavity equator enhances the magnetic field and may cause premature thermal breakdown in the case of superconducting RF cavities; hence, SRF couplers are normally placed in the~beampipe away from the cavity, although some low-field SRF cavities use on-cell couplers.

In normal-conducting cavities, couplers in the beam pipe can either be placed next to the cavity so that the waveguide couples via the iris, or separated from the cavity via a longer, larger diameter circular waveguide in the beam pipe (such that the beam pipe is not cut-off) known as a mode launcher. The~advantage of a mode launcher is that the structure that couples the rectangular waveguide to the circular waveguide can be manufactured separately and connected to the cavity via  a flange, although it takes up more space longitudinally. In many linacs there is a requirement to make the fields as symmetric as possible to avoid a transverse electric or magnetic field on the beam axis which may disrupt the beam. To avoid this two transversely opposing waveguide feeds are often used so that power is fed from both sides.

For SRF couplers the design is complicated by the requirement to minimise the heat conduction between the room-temperature interface and the liquid helium vessel. To minimise the thermal conduction, couplers are often made from steel with a thin coating of copper to minimise ohmic losses on the~RF surfaces. For a given coupler length, it is inefficient to simply have a temperature gradient between the~cold and warm parts; typically there are several stages held at fixed temperatures by cooling with liquid helium at the lowest temperature stage, then helium gas or liquid nitrogen at an intermediate stage in order to minimise the heat deposited at the lowest temperature. Due to the temperature gradient, bellows must be used to allow the coupler to thermally contract when cooling down. 

\index{RF windows}
In addition, to keeping the cavity clean, the coupler will have one or two RF windows which are transparent to RF but which are vacuum tight. The windows will be made from a high-resistivity ceramic -- such as alumina (aluminium oxide) or beryllia (beryllium oxide) -- meaning that the windows have the~problem of charging up if they are struck by electrons; hence, care is taken to avoid any line of sight from the beam to the window. However, the window can still be impacted by electrons due to field emission causing them to charge up. This leads to the possibility of multipactor, vacuum arcs, or flashover -- the~latter where electrons are attracted to the charged ceramic, which on impact produces more secondary electrons, leaving a net positive charge, which in turn are also attracted to the ceramic by the positive charge to give an avalanche. These phenomena can lead to coupler damage, and eventually window metallisation or detuning of the coupler. Multipactor can be avoided in coaxial couplers by providing a DC bias between the inner and outer conductors. Another major cause of window failure is mechanical stress caused by thermal gradients along the window. 

Many coaxial couplers for SRF cavities operating at frequencies above 0.4~GHz will connect to a~rectangular waveguide, and hence a special coupler known as a doorknob is used to transition between the coaxial line and the waveguide. All of the features of an FPC need matching to the RF at the resonant frequency which results in the coupler having a narrow bandwidth.

\subsection {Stub Matching}
Sometimes it is necessary to change the Q of the cavity, this can be done using a 3-stub tuner to match.
If the natural $Q_e$ of the coupling is not matched there is a reflection, however this can be canceled out with another reflection 180 degrees out of phase. Three stubs spaced apart by 0.375 wavelengths can provide any reflection phase/amplitude by varying the stubs insertions. The downside is there is a~standing wave bouncing back and forth between the two reflective surfaces. This creates a standing wave and hence higher peak fields.
Imagine we have two mismatched interfaces to a coaxial line. The~coaxial line is represented by the S matrix and the two reflection coefficients are the stub and the cavity. Lets assume the line is constant impedance so $S_{11}=S_{22}=0$ Lets also assume the line is lossless so $S_{21}=S_{12}=exp(-jkL)$ where k is the propagation constant and L is the distance between them. The~cavity reflection is $(1-\beta)/1+\beta)$ so the stub tuners reflection needs to be the negative of the input impedance to cancel out the cavity reflection.

\begin{equation}
\label{eqn_Stub}
\Gamma_{in} =  \Bigg(  \frac{1- \beta}{1+ \beta} \Bigg) exp(-j 2 k L),
\end{equation}
This technique is commonly used in synchrotrons where the beam-loading can cause large changes in the cavity matching conditions.

\subsection {Coupling with Microphonics}
Microphonics cause the cavity resonant frequency to vary in time by anywhere from 10 Hz to 10 kHz. These frequency deviations also tend to vary sinusoidally in time with frequencies typically at mechanical resonances of the structure. As we have seen before the reflections are dependent on the difference between the drive and natural frequencies. As the detuning angle is proportional to the Q factor it leads to an poor coupling for SRF cavities, which have high Q’s. The required forward power for a cavity that is detuned with respect to the drive frequency is given by
\begin{equation}
\label{eqn_Pfmicro}
P_f =  \frac{V_c^2 (1+\beta)^2}{(8 R \beta} \Big( 1+4 Q_L^2 \frac{\Delta \omega^2}{\omega^2} \Big),
\end{equation}

As can be seen, if the microphonics cause a detuning larger than the cavity bandwidth the power demand to keep the cavity at the required voltage is large. This can lead to amplitude stability problems as the cavity voltage reached for a given drive power will vary in time following the frequency detuning.To avoid this a lower external Q is chosen than the ohmic Q, ie the coupler is not matched. This leads to higher reflections without microphonics but these reflections do not vary in time and hence the RF stability is greatly improved. In addition for a given detuning the external Q factor that leads to the~lowest required drive power for a given voltage shifts down, as both $\beta$ and $Q_L$ are dependant on $Q_e$. This means that operating with a lower Q when detuned actual results in less reflections.

\subsection {On-Crest Beam-Loading}
\index{beam loading}
For the case where the beam is on-crest, cavity behaviour can be described to the first order with some minor modifications to the equations without beam. In this case the beam-loading can be modelled as purely resistive, although it could have a negative resistance for the case of decelerating. The power transferred from the cavity to the beam in the cavity, ignoring the change in the cavity voltage due to the~wake within a single bunch, is approximately given by
\begin{equation}
P_b= V_{acc} I_b,
\end{equation}
where $V_{acc}$ is the accelerating voltage and $I_b$ is the beam current, which must be replaced by the RF source, along with the power to replace ohmic losses in the walls, to maintain the cavity voltage. Looking at the cavity and beam from the RF source, cavity ohmic losses and on-crest beam-loading is indistinguishable, and hence we can define a new coupling factor
\begin{equation}
\beta_b= \frac{P_e}{P_c+P_b}
\end{equation}
and hence the reflected power can be given as
\begin{equation}
P_r= \frac{\omega U}{Q_e} = P_f \Bigg(  \frac{1- \beta_b}{1+ \beta_b} \Bigg)^2
\end{equation}
and the stored energy becomes
\begin{equation}
U_0 = \frac{4 P_f \beta_b^2}{(1+\beta_b)^2} \frac{Q_e}{\omega}.
\end{equation}

\subsection {Off-Crest Beam Loading}
If the beam current is not in phase with the RF voltage, then the beam-loading gains a reactive component, either capacitive or inductive depending on the side of the crest on which the beam arrives. As such, the~beam-loading will change the phase of the RF as well as the amplitude.  Additional RF power will be required due to the reactance, as it will cause reflections at the input coupler. A similar effect will occur if the generator frequency and the cavity frequency are different, with the cavity presenting a reactance to the generator. It is useful to define a detuning angle, $\psi$, given by
\begin{equation}
\tan{\psi}=-2 Q_L \frac{\Delta \omega }{\omega}.
\end{equation}

Considering the power transferred to the beam as well as the reflections due to the change in reactance, or generator detuning, the required RF power, $P_g$, to keep the voltage constant is \cite{wangler}

\begin{equation}
P_g = P_c \frac{(1+\beta)^2}{4 \beta} \frac{1}{\cos^2 \psi} \left[ \left( \cos \phi_s +\frac{V_b \cos \psi}{V_{acc}} \right)^2 +\left( \sin \phi_s +\frac{V_b \sin \psi}{V_{acc}} \right)^2 \right],
\end{equation}
where $\phi_s$ is the phase shift between the cavity voltage and the beam current, $P_c$ is the power required without the beam and $V_b$ is the beam-induced voltage in the cavity given by
\begin{equation}
V_b = \frac{I_b r_s \cos \psi}{1+\beta}.
\end{equation}
The additional required power can be corrected by tuning the cavity to a different resonant frequency to cancel out the beam's reactance. In this case the cavity should be detuned by
\begin{equation}
\tan{\psi}=-2 Q_L \frac{\Delta \omega }{\omega}= - \frac{I R_s \sin{\phi_s}}{V_{acc} (1+ \beta)}.
\end{equation}

\subsection{Matching Travelling-Wave Structures} \index{travelling-wave structure}\index{RF cavity!travelling-wave}

Phase advances that are not integer multiples of 180$^\circ$ result in partially-filled or unfilled cells for standing-wave cavities, as the fields from the forward and backward waves destructively interfere in some cells and constructively in others. This destructive interference can be avoided by using a travelling-wave instead, where the power only flows in a single direction and is absorbed in a load at the other end preventing reflections. The power is fed into the travelling-wave structure via an input coupler and any remaining power is removed at the other end via an output coupler. To avoid standing waves forming inside the travelling wave structure due to reflections at the couplers, each must be carefully matched individually to the structure so that there are no reflections inside the structure. A true travelling wave in vacuum would have a high group velocity requiring too high a power flow to be practical, and the~phase velocity would be greater than the speed of light making synchronisation with a particle beam impossible. To avoid this, the waveguide must be `loaded' to slow the wave down in both group and phase velocity. Whilst this can be done with a uniform dielectric loaded waveguide \cite{DLATHz}, it is more common to load the waveguide with aperture coupled disks, known as a disk-loaded waveguide\index{disk-loaded waveguide}~\cite{TWS}. 

\begin{figure}
\centering
\includegraphics[width=120mm]{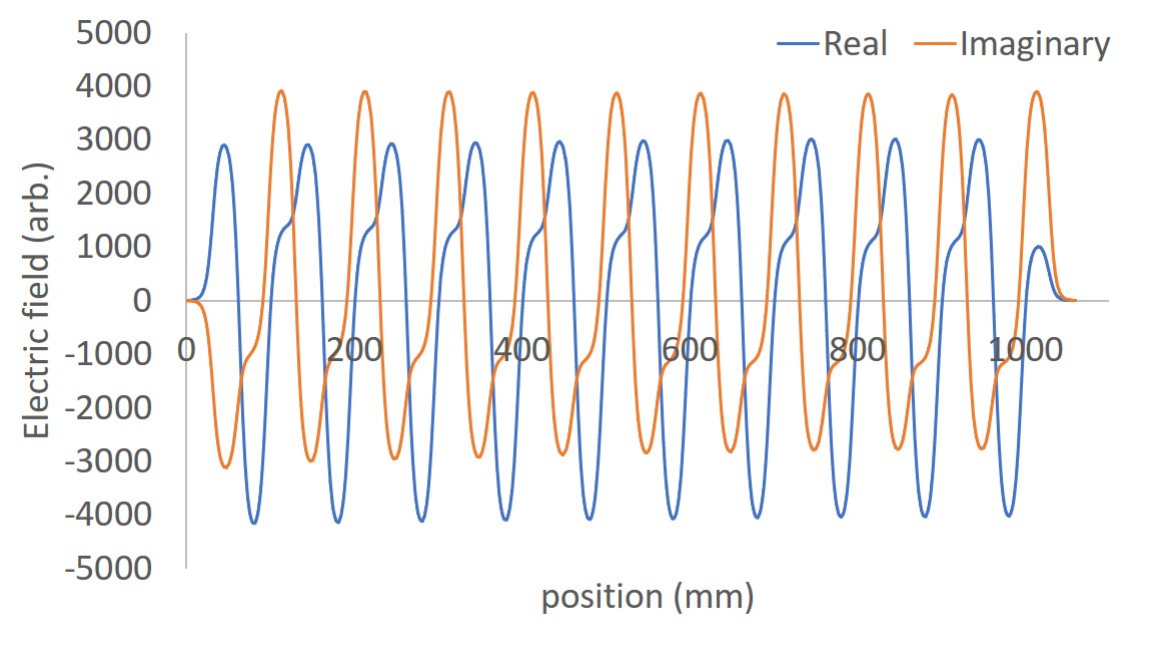}
\caption{\label{fig_TWScomplex} The real and imaginary components of the longitudinal electric field in a $2\pi/3$ travelling-wave structure.}
\end{figure}

\begin{figure}
\centering
\includegraphics[width=140mm]{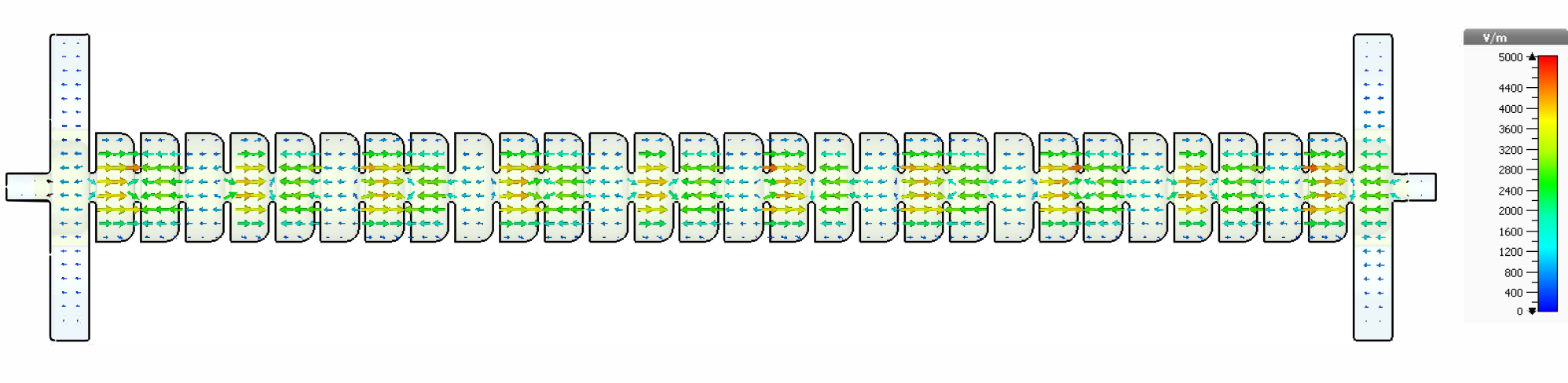}
\caption{\label{fig_TWS} The travelling-wave structure for the AWAKE booster.}
\end{figure}

As the disks are periodic, the wave will be reflected at each disk but will cancel every couple of cells due to the periodicity. As such they are not true travelling waves as each cell will have a longitudinal field variation, making it closer to a chain of standing-wave cells with a phase advance between them. However, the magnitude of the electric field will be identical in each cell and the structure will have a net power flow in one direction unlike a standing-wave structure. Using Floquet's theorem the field in each cell, $E_{cz}$, is identical in each cell other than a phase shift, as shown in Fig.~\ref{fig_TWScomplex} for a 2$\pi$/3 phase advance, and can hence be described using the field profile in a single cell $E_z$ and the phase advance, $\phi_a$ as \cite{Kroll}
\begin{equation}
E_{cz} =E_z (z)[\textrm{exp}(-i \phi_a) + \Gamma \textrm{exp}(i \phi_a)],
\end{equation}
where $\Gamma$ is the reflected wave from the coupler, which is ideally zero for a matched structure. The~travelling-wave structure for the AWAKE booster~\cite{AWAKEboost}, is shown in Fig.~\ref{fig_TWS} showing the cell amplitude at a fixed point in time repeats every three cells, and is hence a $2 \pi / 3$ structure. The amplitude in each cell is constant but there is a phase difference between cells, so at any given point in time, the voltage in each cell will be different. The cell length and the phase advance is chosen such that the beam is always in the cell with the highest voltage. 

Given this, we can evaluate the phase advance and internal reflections inside a structure given the~field in each cell. If we take the sum, $\Sigma$, and difference, $\Delta$, of the fields in the cells either side of a~given cell
\begin{align}
\Sigma &= \frac{(E_z(z+L_{cell})+E_z(z-L_{cell}))}{E_z(z)}, \nonumber\\
\Delta &= \frac{(E_z(z+L_{cell})-E_z(z-L_{cell}))}{E_z(z)}
\end{align}
then the phase advance can be found using
\begin{equation}
\cos(\phi_a)=\frac{\Sigma}{2}
\end{equation}
and the reflected signal can be found from
\begin{equation}
\Gamma=\frac{2 \sin(\phi_a)-i\Delta}{2 \sin(\phi_a)+i\Delta}.
\end{equation}
It should be noted that the internal reflection $\Gamma$ is not the same as $S_{11}$, because if the input and output couplers are identical, the reflection from each coupler will cancel at the input giving $S_{11}=0$ despite there being a reflected wave inside the cavity between the two couplers. For a matched travelling-wave structure, we hence require $S_{11}=\Gamma=0$.

Each cell has a power flow into the cell, $P_w$, power loss in that cell due to ohmic losses or beam loading, and a power flow out of that cell. If the power flow is much larger than the other losses then the~structure has a wider bandwidth, and the cavity behaves like a travelling wave with the filling time of each cell being short compared to the time for the power to flow through the structure. It is this increased bandwidth that makes travelling-wave structures insensitive to imperfections allowing longer structures to be used. They can be at least four times longer than a standing-wave structure. It is possible to load a~short structure so that the group velocity, and hence the power flow, is much lower to increase efficiency. In such cases the individual cells fill slower, and reflections may occur during filling like a hybrid between a travelling- and standing-wave structure~\cite{TERA}.
Due to ohmic losses the power flow decreases along the~structure. The lower the group velocity, the higher the ohmic losses in the cell and hence the power flow will decrease faster along the length of the structure. If the structure is too long, the power will be too low in the end of the structure to achieve any usable gradient hence each structure has a maximum realistic length dependent on group velocity. If the structure is too short, then the power flow at the end of the structure will be large and will be absorbed in an RF load. However, having a lower group velocity also increases the stored energy per cell, and hence the gradient. Therefore for a given structure length, the group velocity should be chosen to maximise the average gradient.

If we consider the matching of the wave flowing from one cell to the next, the cell can be considered as a single cell, strongly coupled to two external couplers (representing the flows to and from the neighbouring cells). In the end cells the couplers must also be matched to apparent external Q factor representing the power flow to the next cell. The Q factor, $Q_f$, related to the power flow between cells is given as,
\begin{equation}
Q_f=\frac{\omega L}{v_g (1-exp(-2 \alpha L))}.
\end{equation}
As the attenuation over a single cell is small, and the cell length is fixed by the phase advance this reduces to
\begin{equation}
Q_e=Q_f=\frac{c \phi_a}{v_g}.
\end{equation}
For a typical group velocity being a few percent of the speed of light, this gives a very low Q compared to that of a standing wave cavity, which leads to a fast filling time and large bandwidth. This is not power loss but power flow so the structure can still be efficient if made sufficiently long.


\newpage
\begin{thebibliography}{99}






\bibitem{Pozar}
D.M Pozar. Microwave Engineering; 3rd ed. Wiley, Hoboken, NJ, 2005.


\bibitem{Padamsee}
H. Padamsee, J Knobloch, T Hays, et al. RF Superconductivity for Accelerators, volume 2011. Wiley
Online Library, 2008.



\bibitem{wangler}
T. P. Wangler. RF Linear Accelerators, Second Edition. Wiley, 2008

\bibitem{DLATHz}
MT Hibberd, AL Healy, DS Lake, V Georgiadis, EJH Smith, OJ Finlay, TH Pacey, JK Jones,
Y Saveliev, DA Walsh, et al. Terahertz-driven acceleration of a relativistic 35 MeV electron beam.
In 2019 44th International Conference on Infrared, Millimeter, and Terahertz Waves (IRMMW-THz),
pages 1–2. IEEE, 2019.

\bibitem{TWS}
C. Nantista, S. Tantawi, and V. Dolgashev. Low-field accelerator structure couplers and design
techniques. Physical Review Special Topics: Accelerators and Beams, 7(7):072001, 2004.
12

\bibitem{Kroll}
N. M. Kroll, C. K. Ng, and D. C. Vier. Applications of time domain simulation to coupler design for
periodic structures. In Proceedings of 20th International Linac Conference, Linac 2000, Monterey,
USA, pages 614–617.


\bibitem{AWAKEboost}
K Pepitone, S Doebert, G Burt, E Chevallay, N Chritin, C Delory, V Fedosseev, Ch Hessler, G Mc-
Monagle, Oznur Mete, et al. The electron accelerator for the AWAKE experiment at CERN. Nuclear
Instruments and Methods in Physics Research Section A: Accelerators, Spectrometers, Detectors and
Associated Equipment, 829:73–75, 2016.


\bibitem{TERA}
S Benedetti, A Grudiev, and A Latina. High gradient linac for proton therapy. Physical Review
Accelerators and Beams, 20(4):040101, 2017.



\end{thebibliography}
\end{document}